\documentclass[preprint,showpacs,preprintnumbers,amsmath,amssymb,nofootinbib]{revtex4}

\usepackage{dcolumn}
\usepackage{bm}
\usepackage{braket}
\usepackage{amsmath}
\usepackage{txfonts}
\usepackage[pdftex]{graphicx,color}
\usepackage{float}

\begin{document}
\preprint{KUNS2632}
\preprint{KOBE-COSMO-16-06}

\title{Oscillating Chiral Tensor Spectrum from Axionic Inflation}
\author{Ippei Obata$^{1}$}
\email{obata@tap.scphys.kyoto-u.ac.jp}
\author{Jiro Soda$^{2}$}
 \email{jiro@phys.sci.kobe-u.ac.jp}

\affiliation{
$^{1}$Department of Physics, Kyoto University, Kyoto, 606-8502, Japan\\
$^{2}$Department of Physics, Kobe University, Kobe, 657-8501, Japan
}

\date{\today}

\begin{abstract}
 We study the axionic inflation with a modulated potential and examine if
the primordial tensor power spectrum exhibits oscillatory feature, which is testable with future space-based gravitational wave experiments such as DECIGO and BBO.
 In the case of the single-field axion monodromy inflation, 
 it turns out that it is difficult to detect the oscillation in the spectrum due to suppression of the sub-Planckian decay constant of axion.
 On the other hand, in the case of aligned chromo-natural inflation where the axion is coupled to a SU(2) gauge field,
 it turns out that the sizable oscillation in the tensor spectrum
 can occur due to the enhancement of chiral gravitational waves sourced by the gauge field.
 We expect that this feature will be a new probe to axion phenomenologies in early universe through the chiral gravitational waves.
\end{abstract}

\pacs{Valid PACS appear here}

\maketitle

\section{Introduction}

 Inflation can be regarded as magnifier which allows us to probe microscopic world, namely, 
high energy physics through the cosmic microwave background (CMB) anisotropies or the large scale structures of the universe.
Indeed, it is important to explore high energy physics such as string theory by means of inflation. 
It is known that there are a lot of scalar fields called axions  and gauge fields in string theory.
It should be noted that an axion is one of the best-motivated candidates of an inflaton since it naturally gives rise to a nearly flat potential
 protected by its shift symmetry \cite{Freese:1990rb}.
 One of the characteristic feature of axionic inflation is that it gets correction of the form of the periodic potential 
due to quantum non-perturbative effects such as instantons.
Specific examples are the axion monodromy mechanism \cite{Silverstein:2008sg} or a kind of aligned natural inflation motivated by the weak gravity conjecture \cite{Kim:2004rp, delaFuente:2014aca}.
 Remarkably, this kind of potential gives the small modulation to  the scalar power spectrum. 
 Thus, the oscillatory feature in the power-spectrum is intimately related to fundamental physics.
 Intriguingly, although the oscillation in the scalar spectrum has been often discussed~\cite{Wang:2002hf, Easther:2013kla},
 the oscillation in the tensor spectrum has been overlooked. 
 This is because, on the CMB scales, the oscillation amplitude in the tensor spectrum is suppressed by several orders of magnitude
 compared to that in scalar spectrum.
 However, it is worth seeking the possibility of this oscillatory signature in the tensor spectrum on the scales probed  by future space-based gravitational wave experiments such as DECIGO \cite{Kawamura:2011zz} and BBO \cite{Crowder:2005nr}.

 In this work, we explore the possibility of generating primordial gravitational waves (PGWs) with oscillatory features in axionic inflation.
 Specifically, we focus on two types of axionic inflations: one is  single-field axion monodromy inflation and the other is a variant of
 inflation driven by the axion coupled to SU(2) gauge field, called aligned chromo-natural inflation \cite{Adshead:2012kp}.
 These two models are quite different from the point of the mechanism of producing tensor spectra.
 In the case of single-field monodromy inflation, the tensor spectrum comes from  vacuum fluctuations.
 We see that it is difficult to detect the oscillatory feature from single-field monodromy inflation 
since the amplitude of oscillation is suppressed by the factor of slow-roll parameters and sub-Planckian decay constant.
 On the other hand, in the case of aligned chromo-natural inflation, the tachyonic growth of
 one helicity mode of the gauge field produces chiral PGWs during inflation \cite{Dimastrogiovanni:2012ew, Adshead:2013qp, Obata:2014loa, Obata:2016tmo}.
 We find that the tensor mode due to particle production of gauge field is sensitive to the modulation of inflaton potential
 and it produces detectable oscillatory feature in the tensor spectrum even for the tiny modulation.
 This feature will open up a new window to physics in early universe through the chiral gravitational waves.

\section{Monodromy inflation}

First, we consider the single-field axionic monodromy inflation.
 The action reads 
\begin{equation}
S=\int dx^4\sqrt{-g}\left[\dfrac{M_p^2}{2}R - \dfrac{1}{2}(\partial_\mu\varphi)^2 - V(\varphi) \right] \ ,
\end{equation}
 where ~$M_p$ ~is Planck mass, ~$R$ ~is Ricci scalar and ~$\varphi$ ~is an axion.
 For the inflaton potential, we use the following form with small modulation generated by the so-called monodromy mechanism 
 in string theory \cite{Silverstein:2008sg}
\begin{eqnarray}
V(\varphi) = V_0(\varphi) + V_\text{mod}(\varphi) 
 = \mu^{4-n}\varphi^n + \Lambda_{\text{mod}}^4\left[1 - \cos\left( \dfrac{\varphi}{f} + \delta \right) \right] \ 
  \label{eq: V1} \ ,
\end{eqnarray}
where ~$V_0$ ~is the bare potential and ~$V_\text{mod}\ll V_0$ is the modulation.
 The parameters ~$\mu, \ n$ ~are constant and ~$\Lambda_{\text{mod}}$ characterizes the size of modulation generated by instanton effects.
 In the cosine potential, ~$f$ ~is a sub-Planckian decay constant of axion and ~$\delta$ ~is a model-dependent phase factor.
Due to this periodic modulation, we can expect the oscillatory feature in the power spectra.

 Let us consider the background dynamics with the modulation.
 The axion obeys the following slow-roll equations
$
3M_p^2H^2 \simeq V \ , \ 3H\dot{\varphi} \simeq - V_\varphi \ .
$
 Note that the dot denotes the derivative with respect to cosmic time ~$t$ ~and ~$V_\varphi \ (V_{\varphi\varphi})$ ~means the 1st (2nd) order derivative of potential with respect to ~$\varphi$ .
 We can define a new variable
~$
b \equiv {\Lambda_{\text{mod}}^4}/{V_{0\varphi} f} 
$
~to parametrizes the strength of the modulation.
 We treat this as a small parameter during inflation in order to ensure that the inflaton does not get trapped.
 Namely, the slope of the bare potential must be larger than that of the modulation
~$
|V_{0\varphi}| > |V_{\text{mod}~\varphi}| 
$ , 
~which implies
$
 |b| \lesssim 1 \label{trap} 
$.
 Defining the bare potential slow-roll parameters
$
\epsilon_{V0} \equiv {M_p^2}/{2}\left({V_{0\varphi}}/{V_0}\right)^2 \ , \  
\eta_{V0} \equiv M_p^2 {V_{0\varphi\varphi}}/{V_0} \ ,
$
the total slow-roll parameters are expressed by
\begin{align}
\epsilon_V &\equiv \dfrac{M_p^2}{2}\left(\dfrac{V_\varphi}{V}\right)^2
\simeq \epsilon_{V0} + 2\epsilon_{V0} b \sin\left( \dfrac{\varphi}{f} + \delta \right) \equiv \epsilon_{V0} + \epsilon_{V1} \ , \\
\eta_V &\equiv M_p^2\dfrac{V_{\varphi\varphi}}{V}= \eta_{V0} + \sqrt{2\epsilon_{V0}}\dfrac{M_p}{f}b \cos\left( \dfrac{\varphi}{f} + \delta \right) \equiv \eta_{V0} + \eta_{V1} \ .
\end{align}
 We can see that ~$\epsilon_{V1}$ ~is suppressed relative to ~$\eta_{V1}$ ~by a factor ~$\sqrt{2\epsilon_{V0}}f/M_p \ll 1$.
 Therefore,  the spectral index of scalar spectrum ~$n_s = 2\eta_V - 6\epsilon_V$ ~is more sensitive to 
 the modulation of the potential compared to that of tensor spectrum ~$n_t = -2\epsilon_V$.
 According to the current CMB  constraint on ~$n_s$ , the amplitude of ~$b$ ~at the pivot scale has an upper bound 
 ~$b f \lesssim 10^{-3}M_p$ ~for ~$f \lesssim 10^{-2}M_p$ \cite{Easther:2013kla}.

 The tensor spectrum with the small oscillation is given by
\begin{eqnarray}
\Delta_h(k) 
 \simeq  \Delta_{h0}(k)\left[ 1 + \sqrt{2\epsilon_{V0}}\dfrac{f}{M_p}b 
\left(1 - \cos\left( \dfrac{\varphi_k}{f} + \delta \right) \right) \right]_{k=aH} \label{eq: tensor} \ ,
\end{eqnarray}
where ~$\Delta_{h0}(k)$ ~is the amplitude of tensor spectrum with no oscillation.
 Note that ~$\varphi_k$ ~is evaluated when a mode with ~$k$ ~exits the horizon.
 Solving slow-roll equations with no modulation, we can approximate ~$\varphi_k$ ~as
\begin{equation}
\varphi_k = \left( \varphi_*^2 - 2n\ln\left(\dfrac{k}{k_*}\right)M_p^2 \right)^{1/2} \simeq \varphi_* - \dfrac{n M_p^2}{\varphi_*}\ln\left(\dfrac{k}{k_*}\right) \ ,
\end{equation}
where ~$\varphi_*$ ~is evaluated at the pivot scale ~$k_*$.
 The second term in the tensor spectrum \eqref{eq: tensor} stems from the modulation of the potential.
 Compared to the scalar spectrum, its amplitude is suppressed by a factor ~$\sqrt{2\epsilon_{V0}}f/M_p \ll 1$.
 Thus, we suspect that its effect is almost invisible with gravitational wave interferometers.

\section{Aligned chromo-natural inflation}

 Next, we consider the effect of modulation on the chromo-natural inflation model.
 The action is as follows:
\begin{eqnarray}
S&=&\int dx^4\sqrt{-g}\left[\dfrac{M_p^2}{2}R - \dfrac{1}{2}(\partial_\mu\varphi)^2 - V(\varphi) \right. \nonumber \\
    && \left. \qquad \qquad \qquad
   - \dfrac{1}{4}F^{a\mu\nu}F^a_{\mu\nu} - \dfrac{1}{4}\lambda\dfrac{\varphi}{f}\tilde{F}^{a\mu\nu}F^a_{\mu\nu} \right] \ ,
\end{eqnarray}
where ~$F^a_{\mu\nu}$ ~is the field strength of SU(2) gauge field
$
F^a_{\mu\nu} = \partial_\mu A^a_\nu - \partial_\nu A^a_\mu + \tilde{g}\epsilon^{abc}A^b_\mu A^c_\nu 
$.
 Note that ~$a$ is the index of SU(2) inner space, ~$\epsilon^{abc}$ ~is Levi-Civita symbol and ~$\tilde{g}$ ~is a gauge coupling constant.
 As a background gauge condition, we choose the temporal gauge ~$A^a_0 = 0$ ~and take an ansatz
\begin{eqnarray}
A^a_i = a(t)Q(t)\delta^a_i \ ,
\end{eqnarray}
which is invariant under the diagonal transformation of the spatial rotation SO(3) and the SU(2)
gauge symmetry.
 Note that ~$a(t)$ ~is scale factor.
 In Chern-Simons term, the dual field strength is defined as
$
\tilde{F}^{a\mu\nu} \equiv \epsilon^{\mu\nu\rho\sigma}F^a_{\rho\sigma} /2
$, 
where ~$\epsilon_{\mu\nu\rho\sigma}$ ~is antisymmetric tensor.
 In this model we suppose the coupling constant ~$\lambda$ 
 ~is much larger than  one.
 Then defining new variables
$
\Lambda \equiv {\lambda}Q/f \ , \ m_Q \equiv {\tilde{g} Q}/{H} 
$,
 if ~$\Lambda^2 \gg 1$ ~and ~$m_Q^2 \gg \Lambda^{-2}$ ~hold, we get the following slow-roll equations
\begin{eqnarray}
Q \simeq Q_{\text{min}} = -\left( \dfrac{f V_\varphi}{3\lambda \tilde{g} H} \right)^{1/3}  \ , \qquad \dfrac{1}{2}\lambda\dfrac{\dot{\varphi}}{f H} \simeq m_Q + \dfrac{1}{m_Q} \label{eq: barsigma} \ .
\end{eqnarray}

 For the axion potential with modulation, we use
\begin{align}
V(\varphi) &= V_0(\varphi) + V_{\text{mod}}(\varphi) \notag \\
 &= \Lambda_0^4 \left[ 1 - \cos \left( \dfrac{\varphi}{f} \right) \right] + \Lambda_{\text{mod}}^4 \left[ 1 - \cos \left( \dfrac{\varphi}{f_{\text{mod}}} + \delta \right) \right] \ .
\end{align}
 Here we defined the energy scale of the bare potential with $\Lambda_0$ and parameters of modulation ~$\Lambda_{\text{mod}}, \ f_{\text{mod}}, \ \delta$.
 Note that above potential form is motivated by instanton effects including a series of higher harmonics terms or KNP mechanism taking into account the weak gravity conjecture \cite{delaFuente:2014aca}.
 Then we get the modified slow-roll equation for ~$Q(t)$ :
\begin{eqnarray}
Q \equiv Q_0 \left( 1 + b\sin\left( \dfrac{\varphi}{f_{\text{mod}}} + \delta \right) \right)^{1/3} \ ,
\end{eqnarray}
where
$
Q_0 \equiv -\left( {f V_{0\varphi}}/{3\lambda \tilde{g} H}\right)^{1/3}  , \  
b \equiv {\Lambda_{\text{mod}}^4}/({V_{0\varphi}f_{\text{mod}}}) \ .
$
 Remarkably, the attractor of the gauge field evolves with oscillation due to the modulation of potential.
 Again, ~$b$ ~must be  small  in order to complete inflation.
 Using the slow roll equation \eqref{eq: barsigma}, the time variation of ~$Q(t)$ ~is approximated by
\begin{equation}
\dfrac{1}{Q}\dfrac{d Q}{H d t} \simeq \dfrac{2}{3\lambda}\dfrac{f}{f_{\text{mod}}}\dfrac{1+m_Q^2}{m_Q}\dfrac{b\cos\left( \dfrac{\varphi}{f_{\text{mod}}} + \delta \right)}{1 + b\sin\left( \dfrac{\varphi}{f_{\text{mod}}} + \delta \right)} \ ,
\end{equation}
where we neglected the time variation of ~$V_{0\varphi}$ ~because of the  slow-roll conditions.
 Thus, we can expect that this modulation gives an oscillatory feature to the tensor spectrum in this model.
 
\begin{figure}[h]
\begin{center}
\includegraphics[width=7cm,height=5cm,keepaspectratio]{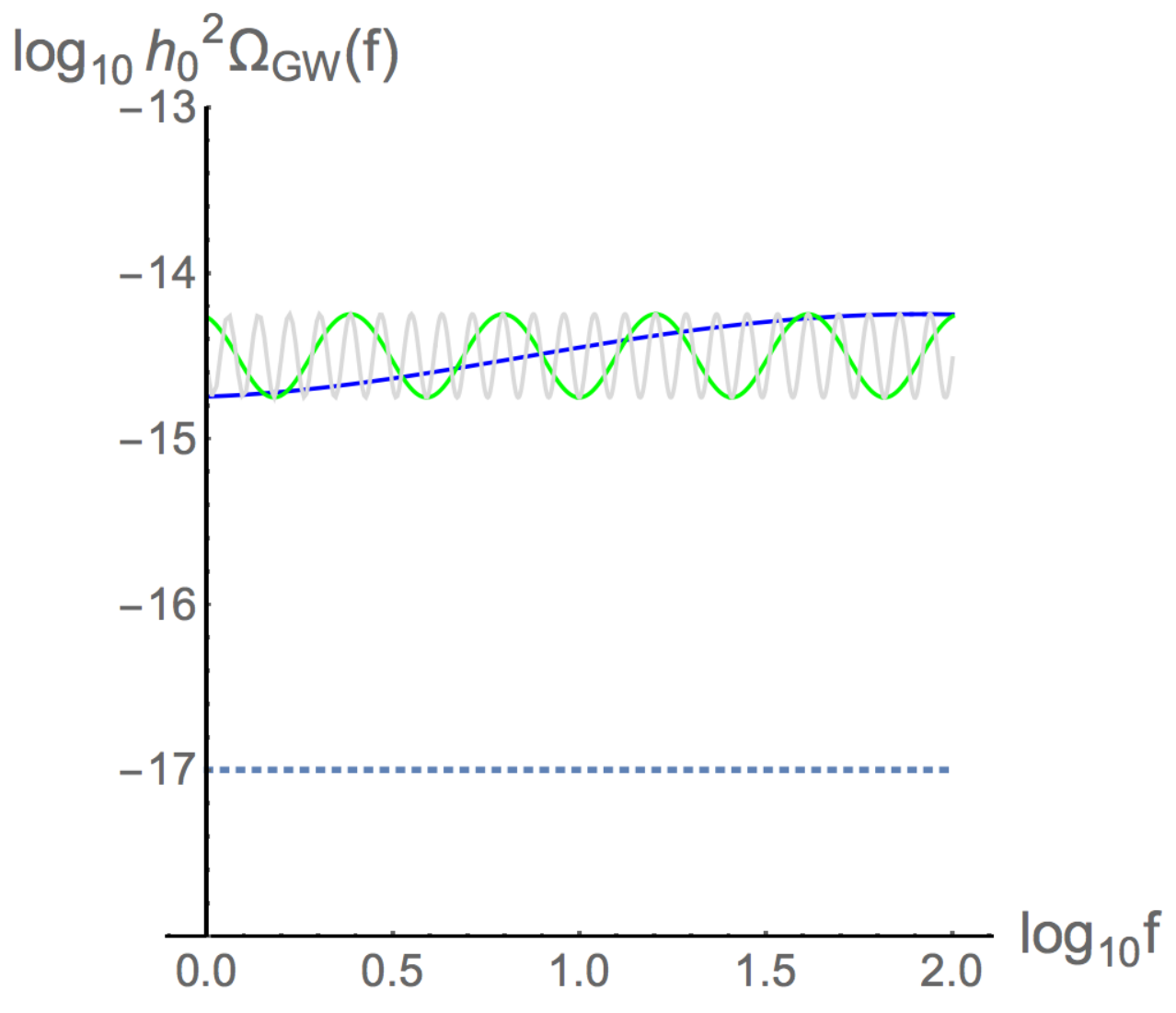}
\end{center}
\caption{The plot of energy density spectrum of chiral gravitational waves.
 The dotted blue line is the spectrum of negative helicity mode.
 On the other hand, that of positive helicity mode with ~$f/\lambda=10^{-3}M_p$ ~(blue line), ~$f/\lambda=10^{-2}M_p$ ~(green line) and ~$f/\lambda=5\times10^{-2}M_p$ ~(light gray line).
 We set ~$(H, \ Q_0, \ m_{Q_0}, \ b) = (10^{-5}M_p, \ 10^{-2}M_p, \ 3, \ 0.1)$ ~in this plot. }
\label{fig: chimodu}
\end{figure}

 As is well known, one of helicity modes of the gauge field experiences tachyonic instability 
around horizon crossing in the presence of the parity violating axion coupling. 
The instability generates parity-violating metric fluctuations at the linear level \cite{Dimastrogiovanni:2012ew, Adshead:2013qp}.
 Thus, the chiral tensor spectra are approximately given by
$
\Delta^{-}_h(k) \simeq \dfrac{H^2}{\pi^2M_p^2} \label{eq: GW-} 
$
and 
\begin{eqnarray}
\Delta^{+}_h(k) \simeq 
\dfrac{H^2}{\pi^2M_p^2}\left[ 1 + 8~|~\chi(Q)|^2 \right] \label{eq: GW+} \ .
\end{eqnarray}
 Here we defined the following enhancement factor \cite{Obata:2016tmo}
\begin{equation}
\chi(Q) \equiv C_2 \dfrac{Q}{M_p} \left( \mathcal{I}_0 - m_{Q}\mathcal{I}_1
 + m_{Q}^2 \mathcal{I}_2 \right) \ ,
\end{equation}
where we have
\begin{eqnarray}
C_2 = -\dfrac{\Gamma(\frac{1}{2} + \mu - \kappa)}{\Gamma(\frac{1}{2} + \mu + \kappa)}(2i)^\kappa (-1)^{\frac{1}{2} + \mu - \kappa} \ , 
\end{eqnarray}
\begin{eqnarray}
    \mathcal{I}_0 &=& \dfrac{ i~\Gamma(-\frac{3}{2} - \mu)\Gamma(-\frac{3}{2} + \mu)}{2} \notag\\
&& \times \ \left(\dfrac{ (\frac{1}{4} - \mu^2- 4\kappa)(\frac{9}{4} - \mu^2) + 8\kappa(1 + \kappa) }{\Gamma(1-\kappa)} \right. \notag \\
&& \left. - \dfrac{(\frac{1}{4} - \mu^2 + 4\kappa)(\frac{9}{4} - \mu^2) - 8\kappa(1 - \kappa)}{\Gamma(\frac{1}{2}-\mu-\kappa)\Gamma(\frac{1}{2}+\mu-\kappa)\Gamma(-\kappa)^{-1}} \right) \ , 
\end{eqnarray}
\begin{eqnarray}
    \mathcal{I}_1 &=& \dfrac{\Gamma(-\frac{1}{2} - \mu)\Gamma(-\frac{1}{2} + \mu)}{2} \notag\\
&& \times\left( \dfrac{\frac{1}{4} - \mu^2 - 2\kappa}{\Gamma(1-\kappa)} + \dfrac{\frac{1}{4} - \mu^2 + 2\kappa}{\Gamma(\frac{1}{2}-\mu-\kappa)\Gamma(\frac{1}{2}+\mu-\kappa)\Gamma(-\kappa)^{-1}} \right) \ , 
\end{eqnarray}
\begin{eqnarray}
   && \mathcal{I}_2 = i~\Gamma(-\tfrac{1}{2} - \mu)\Gamma(-\tfrac{1}{2} + \mu)\notag\\
&&\times\left( \dfrac{1 - 2(1 + \kappa)(\frac{1}{4} - \mu^2)}{\Gamma(-\kappa)} + \dfrac{1 - 2(1 - \kappa)(\frac{1}{4} - \mu^2)}{\Gamma(\frac{1}{2}-\mu-\kappa)\Gamma(\frac{1}{2}+\mu-\kappa)\Gamma(1-\kappa)^{-1}} \right)
\end{eqnarray}
and new variables
~$
\kappa \equiv i (2m_Q + m_Q^{-1})
$
~and 
~$
 \mu^2 \equiv {1}/{4} - 2( m_Q^2 + 1) \ .
$
 Note that this spectrum depends on ~$m_Q$ ~exponentially and is easily enhanced as ~$m_Q$ ~increases.
 So we get a sizable modulation even if ~$b$ ~is small.
 In FIG.\ref{fig: chimodu}, we plotted the energy spectra as a function of the frequency in a DECIGO and BBO range ~$1 ~\text{Hz} < f=k/2\pi < 10^2 ~\text{Hz}$.
 We can see that the spectrum of the positive chirality mode $\Delta^{+}_h(k)$ is enhanced and show oscillation.
 In this plot, we used the following approximate solution of the inflaton derived from \eqref{eq: barsigma}
\begin{equation}
 \varphi_k \simeq \varphi_* + \dfrac{2f}{\lambda}\left(m_{Q_0} + \dfrac{1}{m_{Q_0}}\right)\ln\left(\dfrac{k}{k_*}\right)
\end{equation}
and chose ~$\delta = -\varphi_*/f$.

 Here, one might  worry about the scalar spectrum.
 However, scalar modes of gauge field are stable if ~$m_Q > \sqrt{2}$ ~is satisfied.
 In this region, the curvature power spectrum is approximately given by \cite{Adshead:2013qp}
\begin{equation}
\Delta_{\mathcal{R}} \simeq \dfrac{H^2}{8\pi^2M_p^2\epsilon_H}\dfrac{m_Q^2}{1 + m_Q^2}\bm{Q}(m_Q)^2 \ , \quad \epsilon_H \equiv -\dfrac{\dot{H}}{H^2} \ ,
\end{equation}
where ~$\bm{Q}$ ~is a numerical function which decreases as ~$m_Q$ ~increases.
 Therefore the effect of modulation is not so sensitive for the scalar power spectrum.

\section{detectable parameter estimation}

 Now, we estimate the detectable parameter $b$ from its estimation error by using the correlation analysis with
 gravitational wave interferometers such as DECIGO and BBO.
 Firstly, we consider the data stream ~$V_I$ ~of $I$-th detector in Fourier space which consists of the signal of gravitational waves ~$s_I$ 
~and detector noise ~$n_I$ ,
$
V_I(f) = s_I(f) + n_I(f) \ ,
$
where the index ~$I=\{1,2,1',2'\}$ ~labels each interferometer in two triangle clusters.
 Here, we consider two effectively $L$-shaped interferometers in each cluster whose noises are identical with no correlation each other \cite{Cutler:1997ta}.
 Moreover, we assume that the amplitude of ~$s_I$ ~is much smaller than that of ~$n_I$ .
 The signal is related to the metric tensor modes as
\begin{equation}
s_I(f) = \sum_{A=\pm} \int d\bm{\Omega} h^A(f, \bm{\Omega})F^A_I(\bm{\Omega}) \ ,
\end{equation}
where ~$\bm{\Omega}=(\theta, \phi)$ ~is the direction angle of arrival of the wave and ~$F^A_I$ ~is the pattern function of the $I$-th detector which includes the geometrical information of detectors.
 Note that we set the generic point at the center of star-like detector: ~$\bm{x} = \bm{0}$ .
 For the stochastic background of gravitational waves, the ensemble average of the Fourier amplitude is given by \cite{Maggiore:1999vm}
\begin{align}
&\langle h^A(f, \bm{\Omega})^*h^{A'}(f', \bm{\Omega}') \rangle \notag \\ 
&= \delta_{AA'}\delta(f-f')\dfrac{1}{4\pi}\delta^2(\bm{\Omega} - \bm{\Omega}')\dfrac{3H_0^2}{8\pi^2f^3}\Omega_{GW}(f) \ ,
\end{align}
\begin{align}
&\Omega_{GW}(f) \equiv \dfrac{1}{\rho_c}\dfrac{d\rho_{GW}}{d\ln f} \qquad (~\rho_c = 3M_p^2H_0^2~) \ ,
\end{align}
where
~$H_0 = 100 h_0 ~\text{km}\cdot\text{s}\cdot\text{Mpc}^{-1}$ ~is the present Hubble constant and ~$\Omega_{GW}(f)$ ~is the energy density parameter of the gravitational waves at present.
 Note that ~$h_0 \sim 0.7$ ~is the dimensionless Hubble parameter.
 With the relation $f = k/2\pi$ ,  ~$\Omega_{GW}(f)$ can be expressed by the tensor spectrum as 
$
\Omega_{GW}(f) \simeq 10^{-6}h_0^{-2}\Delta_h(k)
$ \cite{Smith:2005mm}.
 As to the noise variables, they have no correlation between different detectors so that their spectrum can be written as
$
\langle n_I(f)^*n_J(f') \rangle = \delta_{IJ}\delta(f-f') S_I(f) /2 \ .
$
 The noise spectrum of DECIGO and BBO is approximately given by \cite{Yagi:2011wg}
\begin{eqnarray}
S_I^{DECIGO}(f) &&= 6.53\times10^{-49}\left[1 + \left(\dfrac{f}{7.36~\text{Hz}}\right)^2\right] \notag \\
   &&+ 4.45\times10^{-51}\left(\dfrac{f}{1~\text{Hz}}\right)^{-4}\dfrac{1}{1 + \left(\tfrac{f}{7.36~\text{Hz}}\right)^2} \notag \\
 &&+ 4.94\times10^{-52}\left(\dfrac{f}{1~\text{Hz}}\right)^{-4}\text{Hz}^{-1} 
\end{eqnarray}
and
\begin{eqnarray}
S_I^{BBO}(f) &=& 2.00\times10^{-49}\left(\dfrac{f}{1 ~\text{Hz}}\right)^2 + 4.58\times10^{-49} \notag\\
      && + 1.26\times10^{-52}\left(\dfrac{f}{1~\text{Hz}}\right)^{-4}\text{Hz}^{-1} \ .
\end{eqnarray}

 Next, we define the following correlation of data streams
$
\mu_{IJ}(f) \equiv V_I(f)^*V_{J}(f)~\delta f  \ (I \neq J) \ ,
$
where ~$\delta f$ ~is the width of frequency segments.
 Since the noises in different detectors have no correlation, the mean value ~$\langle \mu_{IJ} \rangle$ ~includes only 
the signal of gravitational waves. Then, we get
\begin{equation}
\langle \mu_{IJ}(f) \rangle =  \langle s_I(f)^*s_{J}(f) \rangle \delta f = T_{obs}\dfrac{3H_0^2}{20\pi^2}f^{-3}\gamma_{IJ}\Omega_{GW}(f)~\delta f \ ,
\end{equation}
where ~$T_{obs} = \delta{(f-f)} = \int_{-T_{obs}/2}^{T_{obs}/2}dt \gg \delta f^{-1}$ ~is the time interval of observation.
 Here, we define the overlap reduction function of two detectors as
\begin{equation}
\gamma_{IJ} = \dfrac{5}{2}\sum_{A=\pm}\int \dfrac{d\bm{\Omega}}{4\pi}F^A_I(\bm{\Omega})F^A_{J}(\bm{\Omega}) \ .
\end{equation}
 For isotropic modes, we can set ~$\gamma_{IJ} = 1 ~(~(I, J) = \{(1, 1'), (2,2')\} ~)$ ~for two co-aligned detectors.

 On the other hand, the amount of noise is larger than GW signal, so its variance ~$\sigma_{IJ}^2 = \langle (\mu_{IJ} - \langle \mu_{IJ} \rangle)^2 \rangle$ ~in Fourier mode is approximated by
$
\sigma_{IJ}^2(f) \simeq T_{obs}S_I(f) S_J(f)~\delta f /4
$.
 Therefore, the signal-to-noise ratio ($SNR$) of the stochastic gravitational wave background is given by
\begin{equation}
(SNR)^2 = \sum_{I \neq J}\sum_{f} \dfrac{\langle\mu_{IJ}\rangle^2}{\sigma_{IJ}^2} \simeq \left( \dfrac{3H_0^2}{10\pi^2} \right)^2 T_{\text{obs}}\left[ \sum_{I \neq J}\int_{f_{min}}^{f_{max}}d f \dfrac{\gamma_{IJ}\Omega_{GW}^2}{f^6 S_I S_J} \right] \label{eq: Fi} \ .
\end{equation}
 The summation in \eqref{eq: Fi} is taken with respect to the number of independent detectors of DECIGO and BBO.
 The frequency range is set to ~$f_{min}=0.2~\text{Hz}$ ~and ~$f_{max}=100~\text{Hz}$ ~in order to avoid the binary confusion noise.

 Now, we can discuss the detectability of oscillation.
 We assume that the fiducial parameter of ~$b$ ~is zero.
 Then the amplitude of the parameter estimation error ~$\Delta b$ ~is evaluated by Fisher information quantity ~$\Gamma$ ~given by the inverse of the error correlation \cite{ Seto:2005qy}
\begin{equation}
(\Delta b)^{-2} = \Gamma = \left( \dfrac{3H_0^2}{10\pi^2} \right)^2 T_{\text{obs}}\left[ \sum_{I \neq J}\int_{f_{min}}^{f_{max}}d f \dfrac{\left(\partial_{b}\Omega_{GW}|_{b=0}\right)^2}{f^6 S_I S_J} \right] \label{eq: Fisher} \ .
\end{equation}
 Thus, we can say that oscillation in the spectrum is detectable if $b$  satisfies the inequality
$
b \gtrsim \Gamma^{-{1}/{2}} \ .
$

\subsection{Monodromy inflation}

 We firstly estimate the parameter sensitivity of singe-field version of axion monodromy inflation.
 For simplicity, in our calculation we choose ~$\delta = -\varphi_*/f$ ~and ~$n/\varphi_* = 1/10$ .
 In order to see the oscillation, we need to increase the decay constant of axion.
 However, in order to have oscillation on the inflaton trajectory the decay constant must be smaller than the variation 
of inflaton ~$\Delta\varphi$ ~during inflation
$
{\Delta\varphi}/{f} \gtrsim 1 
$, 
that is,
$
 f \lesssim \sqrt{2\epsilon_V}M_p \sim 10^{-1}M_p \ .
$
 Thus, we have an upper bound on the dacay constant.
 Hence, for the detectability, we obtained the bound
\begin{align}
b &\gtrsim 55\left(\dfrac{T_{obs}}{10\text{yr}}\right)^{-1/2}\left(\dfrac{H}{10^{-5}M_p}\right)^{-2} \quad (\text{DECIGO}) \ , \\
b &\gtrsim 21\left(\dfrac{T_{obs}}{10\text{yr}}\right)^{-1/2}\left(\dfrac{H}{10^{-5}M_p}\right)^{-2} \quad (\text{BBO})
\end{align}
where we set the axion decay constant ~$f = 10^{-1}M_p$.
 Therefore, ~$b$ must be large  in order to get a detectable oscillatory feature, which is not compatible with the requirement of completion of inflation.

\subsection{Aligned chromo-natural inflation}

 Next, we estimate the parameter sensitivity to modulation in aligned chromo-natural inflation.
 Here, we are interested in the parameter region where chiral GWs are sufficiently enhanced.
 In this region, we can approximate ~$\chi$ ~as a real number with a constant complex phase, so that
$
(\partial_b|~\chi(Q)|^2)^2 \simeq | \partial_b ~\chi(Q)^2 |^2
$
is held.
 Using this approximation in \eqref{eq: Fisher}, we calculated the signal to noise ratio and the estimation error for various ~$f/\lambda$ ~values.
 We found that ~$\Delta b$ ~becomes sufficiently small as ~$m_Q$ ~increases.
 This is because we get large ~$SNR$ ~as ~$m_Q$ ~increases and  the tensor spectrum is  exponentially sensitive to $m_Q$.
 More concretely, we get the following lower bound 
\begin{align}
b &\gtrsim 10^{-3.5}\left(\dfrac{T_{obs}}{10\text{yr}}\right)^{-1/2}\left(\dfrac{H}{10^{-5}M_p}\right)^{-2} \quad (\text{DECIGO}) \ , \\
b &\gtrsim 10^{-4}\left(\dfrac{T_{obs}}{10\text{yr}}\right)^{-1/2}\left(\dfrac{H}{10^{-5}M_p}\right)^{-2} \quad (\text{BBO})
\end{align}
where we set ~$Q_0 = f_{\text{mod}} = 10^{-2}M_p$ .
 Therefore, in the case of aligned chromo-natural inflation, we can conclude that 
 the oscillation of gravitational waves is detectable by DECIGO or BBO.


\section{Conclusion}

 We studied the oscillatory feature of tensor spectrum from axionic inflation.
 In the case of single-field monodromy inflation, the modulation in tensor spectrum is too small to be detected by DECIGO or BBO.
 On the other hand, in the case of aligned chromo-natural inflation, we can get the sizable modulation in one helicity mode of tensor perturbation sourced by the gauge field which experienced  the tachyonic instability around horizon crossing during inflation.
 Thus, we found the possibility of producing sizable oscillatory feature in the tensor spectrum of chiral gravitational waves 
from axionic inflations 
when axion couples to the gauge field during inflation.

 In this work, we discussed the tensor spectrum with a sizable oscillation produced by non-Abelian gauge field in chromo-natural inflation. It is known that it is difficult to reconcile the original chromo-natural model  with CMB data 
because it yields too large red scalar spectral index or too much chiral GWs.
 However, it is possible to improve the model so that  chromo-natural inflation occurs in a frequency range higher than nHz and CMB constraints can be satisfied~\cite{Obata:2016tmo}.
 Moreover, we can expect the sizable modulation in the tensor spectrum in the case of Abelian gauge field because one helicity mode of gauge field 
 produces tensor modes at the non-linear level \cite{Sorbo:2011rz}.
 We leave these issues for future work.

\section*{Acknowledgements}

 We would like to thank T.Tanaka and N.Seto for fruitful advice and discussion.
This work was in part supported by JSPS KAKENHI Grant Number 15J01345 and MEXT KAKENHI Grant Number 15H05895.

\end{document}